\documentclass[aps,twocolumn,prl,color,psfig,epsf,nobalancelastpage]{revtex4}
\usepackage{amsmath}
\usepackage{color}
\usepackage{caption}
\usepackage{amsfonts}
\usepackage{epsf}
\usepackage{graphicx}
\usepackage{epstopdf}
\usepackage{ulem}
\usepackage{balance}

\usepackage{bm}

\begin{document}

\title{Transcription-driven genome organization: a model for chromosome structure and the regulation of gene expression tested through simulations}

\author{%
Peter R.~Cook\,$^{1}$,
and Davide Marenduzzo\,$^{2}$%
\footnote{To whom correspondence should be addressed.
Tel: +44 131 6505289; Fax: +44 131 6505902; Email: dmarendu@ph.ed.ac.uk}}

\affiliation{%
$^{1}$Sir William Dunn School of Pathology, University of Oxford, South Parks Road, Oxford, OX1 3RE, 
and
$^{2}$SUPA, School of Physics, University of Edinburgh, Peter Guthrie Tait Road, Edinburgh, EH9 3FD, UK}

\begin{abstract}
 Current models for the folding of the human genome see a hierarchy stretching down from chromosome territories, through A/B compartments and TADs (topologically-associating domains), to contact domains stabilized by cohesin and CTCF. However, molecular mechanisms underlying this folding, and the way folding affects transcriptional activity, remain obscure. Here we review physical principles driving proteins bound to long polymers into clusters surrounded by loops, and present a parsimonious yet comprehensive model for the way the organization determines function. We argue that clusters of active RNA polymerases and their transcription factors are major architectural features; then, contact domains, TADs, and compartments just reflect one or more loops and clusters. We suggest tethering a gene close to a cluster containing appropriate factors -- a transcription factory -- increases the firing frequency, and offer solutions to many current puzzles concerning the actions of enhancers, super-enhancers, boundaries, and eQTLs (expression quantitative trait loci). As a result, the activity of any gene is directly influenced by the activity of other transcription units around it in 3D space, and this is supported by Brownian-dynamics simulations of transcription factors binding to cognate sites on long polymers.
\end{abstract}

\maketitle

\section{Introduction}

Current reviews of DNA folding in interphase human nuclei focus on levels in the hierarchy between looped nucleosomal fibers and chromosome territories~\cite{Dekker2016,Dixon2016}. Hi-C -- a high-throughput variant of chromosome conformation capture (3C) -- provides much of our knowledge in this area. The first Hi-C maps had low resolution ($\sim 1$ Mb), and revealed plaid-like patterns of A (active) and B (inactive) compartments that often contact others of the same type~\cite{LiebermanAiden2009}. Higher-resolution ($\sim 40$ kb) uncovered topologically-associating domains (TADs); intra-TAD contacts were more frequent than inter-TAD ones~\cite{Dixon2012,Nora2012}. Still higher-resolution ($\sim 1$ kbp) gave contact loops delimited by cohesin and CTCF bound to cognate motifs in convergent orientations~\cite{Rao2014},~{{as well as domains not associated with CTCF, called ``ordinary'' or ``compartmental'' domains~\cite{Rao2014,Rowley2017}. [Nomenclature can be confusing, as domains of different types are generally defined using different algorithms.]}}

Despite these advances, critical features of the organization remain obscure. For example, Hi-C still has insufficient resolution to detect many loops seen earlier (Suppl. Note 1). Moreover, most mouse {{domains defined using the Arrowhead algorithm}} persist when CTCF is degraded~\cite{Nora2017} (see also bioRxiv: https://doi.org/10.1101/118737). and many other organisms get by without the protein,~{{(e.g., {\it Caenorhabditis elegans}~\cite{Crane2015}, {\it Neurospora}~\cite{Galazka2016}, budding~\cite{Hsieh2015} and fission yeast~\cite{Mizuguchi2014}, {\it Arabidopsis thaliana}~\cite{Liu2016}, and {\it Caulobacter crescentus}~\cite{Le2016}). Therefore, it seems likely that loops stabilized by CTCF are a recent arrival in evolutionary history. }}

The relationship between structure and function is also obscure~\cite{Dekker2017}. For example, cohesin -- which is a member of a conserved family -- plays an important structural role in stabilizing CTCF loops (Suppl. Note 2), but only a minor functional role in human gene regulation as its degradation affects levels of nascent mRNAs encoded by only $64$ genes~\cite{Rao2017}. Widespread use of vague terms like ``regulatory neighborhood'' and ``context'' reflects this deficit in understanding. Here, we discuss physical principles constraining the system, and describe a parsimonious model where clusters of active RNA polymerases and its transcription factors are major structural organizers -- with contact domains, TADs, and compartments just reflecting this underlying framework. This model naturally explains how genes are regulated, and provides solutions to many current puzzles. 

\enlargethispage{-65.1pt}

\section{Some physical principles}

\subsection{Chromatin mobility}

	Time-lapse imaging of a GFP-tagged gene in a living mammalian cell is consistent with it diffusing for $\sim 1$ minute through a ``corral'' in chromatin, ``jumping'' to a nearby corral the next, and bouncing back to the original one~\cite{Levi2005}. Consequently, a gene explores a volume with a diameter of $\sim 250$ nm in a minute, $\sim 750$ nm in $1$ h, and $\sim 1.4$ $\mu$m in $24$ h~\cite{Lucas2014}; therefore, it inspects only part of one territory in $\sim 24$ h, as a yeast gene -- which diffuses as fast -- ranges throughout its smaller nucleus.

\subsection{Entropic forces}

	Monte Carlo simulations of polymers confined in a sphere uncovered several entropic effects depending solely on excluded volume~\cite{Cook2009,Jun2010}. Flexible thin polymers (``euchromatin'') spontaneously move to the interior, and stiff thick ones (``heterochromatin'') to the periphery -- as seen in human nuclei (Suppl. Fig. S1Ai); ``euchromatin'' loses more configurations (and so entropy) than ``heterochromatin'' when squashed against the lamina, and so ends up internally. Stiff polymers also contact each other more than flexible ones; this favors phase separation and formation of distinct A and B compartments. Additionally, linear polymers intermingle, but looped ones segregate into discrete territories (Suppl. Fig. S1Aii).

\subsection{Ellipsoidal territories and trans contacts}

	Whether a typical human gene diffuses within its own territory and makes cis contacts {{(i.e., involving contacts with the same chromosome)}}, or visits others to make trans ones depends significantly on territory shape. Children who buy M$\&$Ms and Smarties sense ellipsoids pack more tightly than spheres of similar volume; packed ellipsoids also touch more neighbours than spheres (Suppl. Fig. S1B). As territories found in cells and simulations are ellipsoidal, and as much of the volume of ellipsoids is near the surface, genes should make many cis contacts plus some trans ones (Suppl. Fig. S1).

\subsection{Some processes driving looping}

If human chromosomes were a polymer melt in a sphere, two loci $40$ Mbp distant on the genetic map would be $\sim 4$ $\mu$m apart in 3D space and interact as infrequently as loci on different chromosomes. If the two were $10$, $1$ or $0.1$ Mbp apart, they would interact with probabilities of $\sim 2 \times 10^{-5}$, $\sim 5 \times 10^{-4}$, and $\sim 1.5 \times 10^{-2}$, respectively (calculated using a $20$ nm fiber, $50$ bp/nm, and a threshold of $50$ nm for contact detection; see also~\cite{Dekker2016}).  Hi-C shows some contacts occur more frequently; this begs the question -- what drives looping?

	One process is the classical one involving promoter-enhancer contacts~\cite{Rippe2001}. We discuss later that contacting partners are often transcriptionally active. We also use the term ``promoter'' to describe the $5'$ end of both genic and non-genic units, and ``factor'' to include both activators and repressors. Many factors (often bound to polymerases) can bind to DNA and each other (e.g., YY1~\cite{Weintraub2017}). Binding to two cognate sites spaced $10$ kbp apart creates a high local concentration, and -- when two bound factors collide -- dimerization stabilizes a loop if entropic looping costs are not prohibitive (Fig. 1A). Such loops persist as long as factors remain bound (typically $\sim 10$ s).

\begin{figure}[!h]
\begin{center}
\includegraphics[width=0.9\columnwidth]{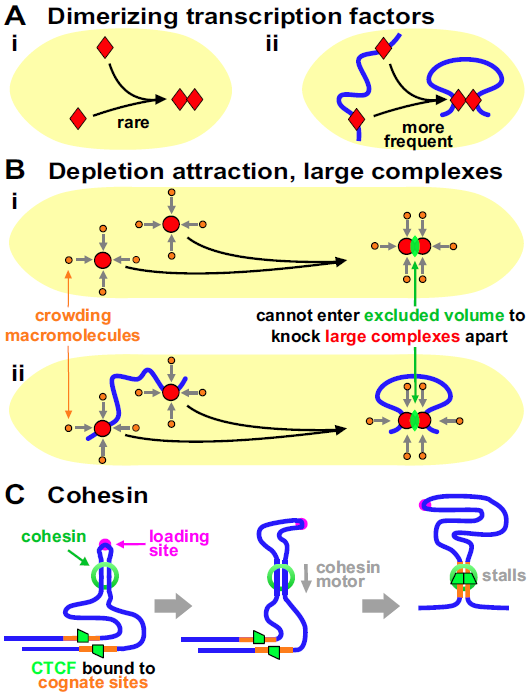}
\end{center}
\caption{Some drivers of looping.
{\bf A.} Dimerizing factors (equilibrium constant $\sim 10^{-7}$ M). {\bf (i)} If present at a typical concentration ($\sim 1$ nM), $<1\%$ factors dimerize. {\bf (ii)} Binding to cognate sites 10 kbp apart on DNA increases local concentrations, and $\sim 67\%$ are now dimers stabilizing loops.
{\bf B.} The depletion attraction. {\bf (i)} In crowded nuclei, small brown molecules (diameter $<5$ nm) bombard (grey arrows) larger red complexes ($5-25$ nm). If large complexes collide, smaller molecules are sterically excluded from the green volume between the two and cannot knock them apart; consequently, small molecules exert a force on opposite sides of larger complexes keeping them together. {\bf (ii)} If large complexes are bound to DNA, this force stabilizes a loop.
{\bf C.} Cohesin. After loading, a cohesin ring embraces two fibers to stabilize a mini loop; this loop enlarges as the ring uses an inbuilt motor to move down the fiber until stalled by CTCF bound to convergent sites.}
\label{StrucBasisFig1}
\end{figure}

Another mechanism -- the ``depletion attraction'' -- is non-specific. It originates from the increase in entropy of macromolecules in a crowded cell when large complexes come together (Fig. 1Bi~\cite{Marenduzzo2006}). Modeling indicates this attraction can cluster bound polymerases and stabilize loops (Fig. 1Bii) that persist for as long as polymerases remain bound (i.e., seconds to hours; below). 

	A third mechanism involves cohesin -- a ring-like complex that clips on to a fiber like a carabiner on a climber's rope. In Hi-C maps, many human domains are contained in loops apparently delimited by CTCF bound to cognate sites in convergent orientations~\cite{Rao2014}. Such ``contact loops'' -- many with contour lengths of $>1$ Mbp -- are thought to arise as follows. A cohesin ring binds at a ``loading site'' to form a tiny loop, this loop enlarges as an in-built motor translocates the ring down the fiber, and enlargement ceases when CTCF bound to convergent sites blocks further extrusion (Fig. 1C~\cite{Sanborn2015,Fudenberg2016}). {{This is known as the ``loop-extrusion model''. We note that other mechanisms could enlarge such loops (including one not involving a motor; Suppl. Note 2), and that loop extrusion (by whatever mechanism) and its blocking by convergent CTCF sites can be readily incorporated into the model that follows.}}

\subsection{A transcription-factor model}

We now review results of simulations involving what we will call the ``transcription-factor model''. This incorporates the few assumptions implicit in the classical model illustrated in Figure 1A: spheres (``factors'') bind to selected beads in a string (``cognate sites'' on ``chromatin fibers'') to form molecular bridges stabilizing loops~\cite{Barbieri2012,Brackley2013,Brackley2016,Bianco2017,Haddad2017}. This superficially simple model yields several unexpected results. 

First, and extraordinarily, bound factors cluster spontaneously in the absence of any specified DNA-DNA or protein-protein interactions (Fig. 2A~\cite{Brackley2013}). This clustering requires bi- or multi-valency (so factors can bridge different regions and make loops) plus reversible binding (otherwise the system does not evolve), and it occurs robustly with respect to changes in DNA-protein affinity and factor number. The process driving it was dubbed the ``bridging-induced attraction''~\cite{Brackley2013}.  We stress this attraction occurs spontaneously without the need to specify any additional forces between one bead and another, or between one protein and another. 

\begin{figure*}[!h]
\begin{center}
\includegraphics[width=\columnwidth]{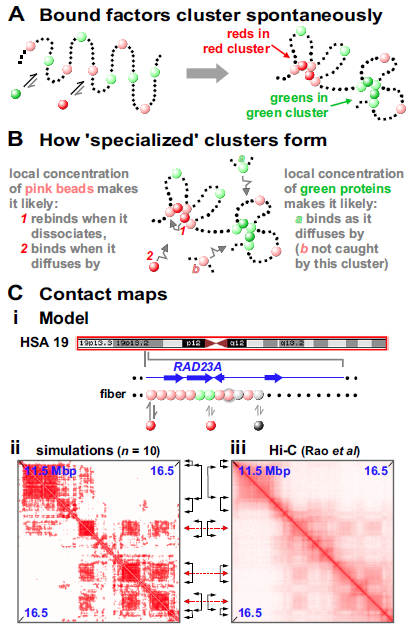}
\end{center}
\caption{{{A process driving the spontaneous clustering of multivalent factors (a.k.a., the ``bridging-induced attraction'').}} 
{\bf A.} Overview of one Brownian-dynamics simulation. Red and green ``factors'' (colored spheres) bind reversibly to ``chromatin'' (a string of beads); red factors bind only to pink beads, green factors only to light-green ones (non-binding beads shown as black dots). Bound factors spontaneously cluster -- red with red, and green with green -- despite any specified interactions between proteins or between beads.
{\bf B.} Explanation. Local concentrations create positive-feedback loops driving growth of nascent clusters; bound factors and binding beads rarely escape, and additional factors/beads are caught as they diffuse by. Red and green clusters are inevitably separate in 3D space because their cognate binding sites are separate in 1D sequence space. Cluster growth is limited by entropic costs of crowding together ever-more loops.
{\bf C.} 
Comparison of contact maps obtained from 10 simulations~\cite{Brackley2016} and Hi-C~\cite{Rao2014}. (i) The model. The whole of chromosome 19 (red box) in GM12878 cells was simulated, and the zoom shows the region around {\it RAD23A}, which is active in these cells. Each bead in the fiber is colored according to whether the corresponding region is transcriptionally highly active (pink), weakly active (green), or silent (grey) on the Broad ChromHMM track on the UCSC browser; one bead carries both active and silent marks and so bears two colors. Pink (activating) and black (repressing) factors bind to cognate beads as indicated (the doubly-colored bead binds both factors); all other beads (black dots) are non-binding. (ii, iii). Contact maps are similar. Black double-headed arrows: limits of prominent TADs on diagonal. Red double-headed arrows: centers of off-diagonal blocks marking compartments.}
\end{figure*}

{{The basic mechanism yielding clustering is a simple positive feedback loop which works as sketched in Figures~2A,B. First, proteins bind to chromatin (Fig.~2A). Then, once a bridge forms, the local density of binding sites (e.g., pink spheres in Fig.~2A) inevitably increases. This attracts further factors from the soluble pool (like {\it 2} in Fig.~2B): their binding further increases the local chromatin concentration (through bridging) creating a virtuous cycle which repeats. This triggers the self-assembly of stable protein clusters, where growth is eventually limited by entropic crowding costs~\cite{Brackley2016}.}} 
{{Several factors cluster in nuclei (e.g., Sox2 in living mouse cells~\cite{Liu2014}) and the bridging-induced attraction provides a simple and general explanation for this phenomenon.}} 


{{This process drives local phase separation of polymerases and factors, and so naturally explains how super-enhancer (SE) clusters form (Suppl. Fig. S2Ai~\cite{Hnisz2017}). This generic tendency to cluster will be augmented by specific protein--protein and DNA--protein interactions, with their balance determining whether protein or DNA lies at the core. Similarly, the same process -- this time augmented by HP1, a multivalent protein that staples together histones carrying certain modifications -- could drive phase separation and compaction of inactive heterochromatin (Suppl. Fig. S2B~\cite{Larson2017,Strom2017}). }}

\subsection{Creating stable clusters of different types, TADs, and compartments}

	This transcription-factor model yields a second remarkable result: red and green factors binding to distinct sites on the string self-assemble into distinct clusters containing only red factors or only green ones (Fig. 2A~\cite{Brackley2016}). This has a simple basis: the model specifies that red and green binding sites are separate in 1D sequence space (as they are {\it in vivo}), so they are inevitably in different places in 3D space (Fig. 2B). 


	A third result is that clusters and loops self-assemble into ``TADs'' and ``A/B compartments''~\cite{Barbieri2012,Brackley2013,Brackley2016}. Thus, if chromosome 19 in human GM12878 cells is modeled as a string of beads colored according to whether corresponding regions are active or inactive, binding of just red and black spheres (``activators'' and ``repressors'') yields contact maps much like Hi-C ones (Fig. 2C). As neither TADs, compartments, nor experimental Hi-C data are used as inputs, this points to polymerases and their factors driving the organization without the need to invoke roles for higher-order features (see also~\cite{Rowley2017}). We suggest TADs arise solely by aggregation of pre-existing loops/clusters (note that degradation of cohesin or its loader induces TAD disappearance and the emergence of complex sub-structures, as A/B compartments persist and become more prominent~\cite{Rao2017,Schwarzer2017}).

The simple transcription-factor model has been extended to explain how pre-existing red clusters can evolve into green clusters, or persist for hours as individual factors exchange with the soluble pool in seconds -- as in photo-bleaching experiments (Suppl. Fig. S3A,B~\cite{Brackley2016,Brackley2017a}). Additionally, introducing ``bookmarking'' factors that bind selected beads (genomic sequences), as well as ``writers'' that ``mark'' chromatin beads and ``readers'' which bind beads with specific marks, can create local ``epigenetic states'' and epigenetic domains (e.g., domains of red and green marks, representing for instance active or inactive histone modifications). Such domains spontaneously establish around bookmarks, and are stably inherited through ``semi-conservative replication'', when half of the marks are erased (and/or some of the bookmarks are lost due to dilution~\cite{Michieletto2016,Michieletto2017}; Suppl. Fig. S3C).  

\clearpage

\section{A parsimonious model: clusters of polymerases and factors}

These physical principles lead naturally to a model in which a central architectural feature is a cluster of active polymerases/factors surrounded by loops -- a ``transcription factory''. A factory was defined as a site containing $\ge 2$ polymerases active on $\ge 2$ templates, just to distinguish it from cases where $\ge 2$ enzymes are active on one (Fig. 3A~\cite{Rieder2012,Papantonis2013}). Much as car factories contain high local concentrations of parts required to make cars efficiently, these factories contain machinery that acts through the law of mass action to drive efficient RNA production. For RNA polymerase II in HeLa, the concentration in a factory (i.e., $\sim 1$ mM) is $\sim 1,000$-fold higher than the soluble pool; consequently, essentially all transcription occurs in factories (Suppl. Note 3; Suppl. Note 4 describes some properties of factories).

\begin{figure}[!h]
\begin{center}
\includegraphics[width=0.75\columnwidth]{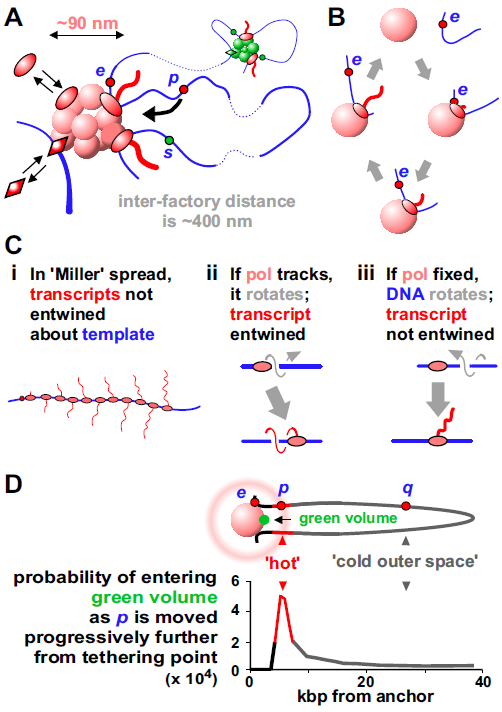}
\end{center}
\caption{Transcription factories in human cells.
 {\bf A.} Clusters organize loops stabilized by polymerases (ovals) and factors (lozenges). There are $\sim 16$ loops per factory, but only a few are shown. Red and green factories specialize in transcribing different gene sets. Promoters tend to be transcribed in factories of the same color (because they are rich in appropriate factors); here, {\it p} and {\it s} can often visit the pink factory, but only {\it p} is likely to initiate there. 
{\bf B.} A transcription cycle. Promoter {\it e} collides with a polymerase in the factory (shown as a solid sphere from now on), initiates, and the fixed polymerase reels in the template as it extrudes a transcript; the template detaches on termination.
{\bf C.} ``Miller'' spreads. {\bf (i)} A Christmas tree. {\bf (ii)} If the polymerase tracks, it rotates about the template once for every $10$-bp transcribed to give an entwined transcript. {\bf (iii)} If immobile, the template rotates and the transcript is not entwined. Topoisomerases remove twin domains of supercoiling in both (ii) and (iii)~\cite{Ahmed2017}.
{\bf D.} Tether length affects how often a promoter visits a factory. Top: a $77$-kbp loop tethered to a $75$-nm sphere; intuition suggests {\it p} visits the green volume more than {\it q}. Bottom: results of Monte-Carlo simulations confirm this intuition. Adapted from~\cite{Bon2006} with permission.}
\end{figure}

In all models, a gene only becomes active if appropriate polymerases (i.e., I, II, or III) and factors are present; in this one, there are $3$ more requirements. First, active polymerases are transiently immobile when active; they reel in their templates as they extrude their transcripts (Fig. 3B). This contrasts with the traditional view where they track like locomotives down templates. Arguably, the best (perhaps only) evidence supporting the traditional view comes from iconic images of ``Christmas trees''; a 3D structure is spread in 2D, and imaged in an electron microscope -- polymerases are caught in the act of making RNA (Fig. 3Ci). However, polymerases moving along helical templates generate entwined transcripts (Fig. 3Cii), but these transcripts appear as un-entwined ``branches'' in ``Christmas trees''. How could such structures arise? As transcription requires lateral and rotational movement along/around the helix, we suggest templates move (not polymerases) to give un-entwined transcripts (Fig. 3Ciii). Consequently, these images provide strong evidence against the traditional model, not for it (see also Suppl. Note 5, Suppl. Fig. S4). 

	Second, to initiate, a {{promoter must have a high probability of colliding with a polymerase, and -- as the highest polymerase concentractions are found in/around factories -- this means the enzyme must first diffuse into/near a factory. [We remain agnostic as to the order with which promoter, polymerase, factors and factory bind to each other, and note that the participants in nucleotide excision repair -- a process arguably better understood than transcription~\cite{Dinant2009} -- are not assembled one after the other; instead the productive complex forms once all participants collide simultaneously into each other.]}} In Figure 3D, intuition suggests {\it p} often visits the nearby green volume, whereas {\it q} mainly roams ``outer space''; simulations and experiments confirm this~\cite{Bon2006,Larkin2013}. {{Consequently, active genes tend to be tethered close to a factory, and inactive genes further away.}} Promoter-factory distances also seem to remain constant as nuclear volume changes; when mouse ES cells differentiate and their nuclei become two-fold larger or two-fold smaller, experiments show the system spontaneously adapts to ensure these distances remain roughly constant, and new simulations confirm this (Suppl. Fig. S6; Suppl. Note 6).

	\clearpage

	Third, there are different types of factory (red and green clusters in Fig. 3A), and a gene must visit an appropriate one to initiate. Just as some car factories make Toyotas and others Teslas, different factories specialize in transcribing different sets of genes. For example, distinct ``ER$\alpha$'', ``KLF1'', and ``NF$\kappa$B'' factories specialize in transcribing genes involved in the estrogen response, globin production, and inflammation, respectively~\cite{Fullwood2009,Schoenfelder2010,Papantonis2012}.

        {{These three principles combine to ensure the structure is probabilistic and dynamic, with current shape depending on past and present environments. For example, as {\it e} in Figure 3D is transcribed, loop length changes continuously. And when {\it e} terminates, it dissociates; then, its diffusional path may take it back to the same factory where it may (or may not) re-initiate to reform a loop. Alternatively, {\it e} may spend some time diffusing through outer space before rebinding to the same or a different factory. Consequently, as factors and polymerase bind and dissociate, factories morph, loops appear and disappear -- and the looping pattern of every chromosomal segment changes from moment to moment. Then, it is unlikely the 3D structure of any chromosome is like that of its homolog, either in the same cell or any other cell in a clonal population.}}

        {{These physical principles also lead naturally to an explanation of how genes become inactive. Thus, {\it q} in Figure 3Di is inactive because it lies far away from an appropriate factory and is unlikely to collide with a polymerase there. We speculate that inactivity results in histone modifications that thicken the fiber, so entropic effects collapse it with other heterochromatic fibers into B compartments and the nuclear periphery (as in Suppl. Fig. S1Ai).}}

\section{Some difficult-to-explain observations}

We now describe results easily explained by this model, but difficult or impossible to explain by others without additional complicated assumptions (see also Suppl. Note 7). 

\subsection{Most contacts are between active transcription units}

	Contacts seen by 3C-based approaches often involve active promoters and enhancers; for example, FIRES (frequently-interacting regions) in 14 different human tissues and 7 human cell lines are usually active enhancers~\cite{Schmitt2016a}. 
Similarly, contacts detected by an independent method -- genome architecture mapping -- again involve enhancers and/or genic transcription start/end sites~\cite{Beagrie2017}. Why should active sequences lie together? As factories nucleate local concentrations of active units, we expect promoters and enhancers to dominate contact lists.
 
	While 3C focuses on contacts between two DNA sequences, the ligation involved can join $>2$ together ($24$ is the current record), and these again generally encode active sequences~\cite{Ay2015,Olivares2016}. Why do so many active sequences contact each other? We expect to see co-ligations involving some/all of the many 
anchors in a typical factory.

	Early studies also point to a correlation between transcription and structure. For example, switching on/off many mammalian genes correlates with their attachment/detachment~\cite{Papantonis2013}. What underlies this? Our model requires that units must attach before they can be transcribed. 

\subsection{Frequencies of cis and trans contacts}


        Cis Hi-C contacts fall off rapidly with increasing genetic distance, whereas trans ones are so rare they are often treated as background. However, ChIA-PET yields more trans than cis contacts when active sequences are selected by pulling down ER$\alpha$ or polymerase II~\cite{Fullwood2009,Papantonis2012}. Our model again predicts this -- active genes on different chromosomes are often co-transcribed in the same specialized factory (as genes diffuse out of one ellipsoidal territory into another).  


	In addition, cis:trans ratios can change rapidly, and we explain this by reference to ``NF$\kappa$B'' factories~\cite{Papantonis2012} (see also Suppl. Note S3 and Suppl. Fig. S5A). TNF$\alpha$ induces phosphorylation of NF$\kappa$B, nuclear import of phospho-NF$\kappa$B, and transcriptional initiation of many inflammatory genes including {\it SAMD4A}. Before induction, the {\it SAMD4A} promoter makes only a few local cis contacts (shown by 4C and ChIA-PET applied with a ``pull-down'' of polymerase II); it spends most time roaming ``outer space'' making a few chance contacts with nearby segments of its own loop, and -- if it visits a factory -- it cannot initiate in the absence of phospho-NF$\kappa$B. But once phospho-NF$\kappa$B appears (10 min after adding TNF$\alpha$), it initiates. Then, NF$\kappa$B binding sites in {\it SAMD4A} become tethered to the factory, these bind phospho-NF$\kappa$B, exchange of the factor increases the local concentration, and this increases the chances that other inflammatory genes initiate when they pass by. And once they do, this creates a virtuous cycle; as more inflammatory genes initiate, more NF$\kappa$B binding sites become tethered to the factory, the local NF$\kappa$B concentration rises, this further increases the chances that passing responsive genes initiate, and the factory evolves into one specializing in transcribing inflammatory genes. As a result, the rapid concentration of inflammatory genes around the resulting ``NF$\kappa$B'' factory yields the rapid increase in cis and trans contacts between them seen by 3C-based methods and RNA-FISH~\cite{Papantonis2012}.

\subsection{TADs exist at all scales}

	Intra- and inter-TAD contact frequencies differ only $\sim 2$-fold; therefore, it is unsurprising that TAD calling depends on which algorithm is used, and the resolution achieved~\cite{Schmitt2016b,Dali2017,Forcato2017,Zhan2017}. However, it is surprising that TADs become more elusive as algorithms and resolution improve. For example, CaTCH (Caller of Topological Chromosomal Hierarchies) identifies a continuous spectrum of domains covering all scales; TADs do not stand out as distinct structures at any level in the hierarchy~\cite{Zhan2017}. Moreover, TADs are sometimes invisible in single-cell data~\cite{Flyamer2017,Stevens2017}, and -- if detected -- their borders weaken as cells progress through G1 into S phase~\cite{Nagano2017}. In our model, TADs do not exist as distinct entities representing anything other than one or more loops around one or more factories. [TADs are said to be major architectural features~{{because they are invariant between cell types~\cite{Dixon2012,Nora2012} and highly conserved~\cite{Harmston2017}. However, there are always slight differences between cell types that could reflect slight differences in expression profile, and the conservation could just reflect}} the conserved transcriptional pattern encoded by the underlying DNA sequence.]

\subsection{The relationship between TADs and transcription}

Various studies address this issue, and give conflicting results. For example, in mouse neural progenitor cells, one of the two X chromosomes is moderately compacted and largely inactive. Inactive regions do not assemble into A/B compartments or TADs, unlike active ones. Moreover, in different clones, different regions in the inactive X escape inactivation, and these form TADs~\cite{Giorgetti2016}. Here, structure and activity are tightly correlated (in accord with our model). Similarly, inhibiting transcription in the fly leads to a general reorganization of TAD structure, and a weakening of border strength~\cite{Li2015}. 

Another study points to some TADs appearing even though transcription is inhibited~\cite{Hug2017}. After fertilization, the zygotic nucleus in the fly egg is transcriptionally inactive. As the embryo divides, zygotic genome activation occurs so that by nuclear cycle $8$ (nc8), $\sim 180$ genes are active, and these seem to nucleate a few TADs detected at nc12 (so transcriptional onset and the appearance of {{loops/TADs}} correlate -- again in accord with our model). As more genes become active at nc13, $3$-fold more TADs develop by nc14, and polymerase II plus Zelda (a zinc-finger transcription factor) are at boundaries (again a positive correlation). If transcriptional inhibitors are injected into embryos before nc8, boundaries and TADs seen at nc14 are less prominent, but some TADs still develop (implying {{loops/TADs}} appear independently of transcription, which is inconsistent with our model). However, interpretation is complicated. Although inhibitors reduce levels of $5$ mRNAs already being expressed, they only slightly affect levels of polymerase II bound at the $5'$ end of genes expressed at nc14; this indicates that inhibition is inefficient, so it remains possible that the remaining transcription stabilizes the {{loops/TADs}} seen.
 
Studies on mouse eggs and embryos also provide conflicting data. Thus, activity is lost as oocytes mature, and TADs plus A/B compartments disappear~\cite{Du2017,Flyamer2017,Ke2017}; therefore, loss of structure and activity again correlate (consistent with our model). After fertilization, the zygote contains two nuclei with different conformations; both contain TADs, but the maternal one lacks A/B compartments. Then, as transcription begins, TADs appear (again a positive correlation), but $\alpha$-amanitin (a transcriptional inhibitor) does not prevent this~\cite{Du2017,Ke2017} -- which is inconsistent with our model. However, interpretation is again complicated: $\alpha$-amanitin acts notoriously slowly~\cite{Bensaude2011}, and inhibition was demonstrated indirectly (levels of steady-state poly(A)$^{+}$ RNA fall, but reduction of intronic RNA would be a more direct indicator of inhibition).
 
{{	
Data from zebrafish make unified interpretation even more difficult. In contrast to some cases cited earlier, TADs and compartments exist before zygotic gene activation, and many of each are lost when transcription begins~\cite{Kaaij2018}. Clearly, TAD-centric models will find it difficult to explain such conflicting data. In ours, TADs are not major architectural features determining function; they just reflect the underlying network of loops, and -- even if all polymerases are inactive -- bound factors can still stabilize some loops (and so TADs).
}}

\subsection{Enhancers and super-enhancers}

	Enhancers are important regulatory motifs, but there remains little agreement on how they work~\cite{Long2016}. They were originally defined as motifs stimulating firing of genic promoters when inserted in either orientation upstream or downstream. However, their molecular marks are so like those of their targets~\cite{Kim2015} that FANTOM5 now defines them solely as promoters firing to yield eRNAs (enhancer RNAs) rather than mRNAs~\cite{Andersson2014}. Then, is it eRNA production or some role of the eRNA product that underlies function? Studies of the {\it Sfmbt2} enhancer in mouse ES cells indicates it is the former~\cite{Engreitz2016}. Thus, deleting the eRNA promoter (but not downstream sequences) impairs enhancer activity; this points to the promoter being required. Moreover, inserting a poly(A) site just 40 bp down-stream of the eRNA promoter abolishes enhancer activity, and amounts of polymerase on the enhancer (and enhancer activity) increase as the insert is moved progressively $3'$; this points to a reduction in transcription correlating with reduced enhancer activity. 

	Our model suggests a simple mechanism for enhancer function: transcription of {\it e} in Figure 4Ai ensures {\it p} is tethered close to an appropriate factory. In other words, {\it e} is an enhancer of {\it p} because close tethering increases the probability that {\it p} collides with a polymerase in the factory (and so often initiates). The model also explains how enhancers can act over such great distances (Suppl. Fig. S5B,C). Thus, a typical factory in a human cell is associated with $\sim 10$ loops each with an average contour length of $\sim 86$ kbp (Suppl. Note 1), so an enhancer anchored to it can (indirectly) tether a target promoter in any one of these other loops to the same factory. As we will see, enhancers can act over even greater distances to tether targets in a nuclear region containing an appropriate factory.

\begin{figure}[!h]
\begin{center}
\includegraphics[width=\columnwidth]{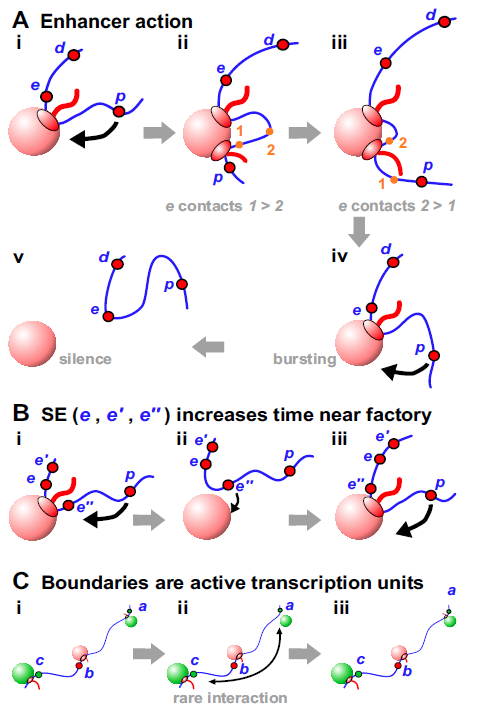}
\end{center}
\caption{Enhancers and boundaries.
{\bf A.} Enhancer action. {\bf (i)} {\it p} is tethered by enhancer {\it e} close to a factory -- so {\it p} is likely to collide with the factory. {\bf (ii)} {\it p} has initiated, and the polymerase is about to transcribe {\it 1}. {\bf (iii)} The same polymerase will now transcribe {\it 2}; then, {\it e}-{\it p} contacts apparently track with the polymerase away from {\it p}. Both polymerases now terminate, {\it e} and {\it p} detach, and {\it e} reinitiates. {\bf (iv)} As {\it p} is still tethered close to the factory, it is likely to initiate again and continue the transcriptional burst. {\bf (v)} Both polymerases have terminated, and the fiber has diffused away from the factory; both {\it e} and {\it p} enter a silent period, as both are far from the factory.
{\bf B.} SEs increase the time {\it p} is close to a factory. {\bf (i)} The structure is as Ai, but now the enhancer contains $3$ promoters; as before, {\it p} is tethered close to a factory and likely to initiate. {\bf (ii)} The polymerase transcribing {\it e} has terminated; as there are $3$ SE promoters, there is a $3$-fold higher chance one will collide with the factory (here {\it e''}) compared to A. {\bf (iii)} {\it e''} has initiated, so {\it p} remains closely-tethered for longer and likely to initiate more often than in A.
{\bf C.} Boundaries. {\bf (i)} {\it a}, {\it b}, and {\it c} have initiated in different factories. {\bf (ii)} {\it a} has terminated, and is more likely to visit the upper green factory compared to the distant lower one. {\bf (iii)} {\it a} has re-initiated in the nearby green factory. We call {\it b} a boundary because it apparently prevents {\it a} from contacting {\it c}.
}
\end{figure}

This model provides solutions to many conundrums associated with enhancers, including: (i) Enhancer activity depends on contact with its target promoter~\cite{Deng2014,Levine2014}. We suggest the two often share a factory, and so are often in contact. (ii) Enhancers can act on two targets simultaneously, and coordinate their firing~\cite{Fukaya2016,Muerdter2016} -- impossible according to classical models. In Figure 4Ai, {\it e} acts on both {\it d} and {\it p}, and it is easy to imagine that {\it d} and {\it p} initiate coordinately because the two polymerases involved sit side-by-side in the same factory. (iii) Promoters of protein-coding genes are often enhancers of other protein-coding genes~\cite{Engreitz2016,Dao2017,Diao2017}. In our model, {\it e} is an enhancer irrespective of whether it encodes an mRNA or eRNA. (iv) Enhancers act both promiscuously and selectively. They interact with many other enhancers and targets~\cite{Javierre2016,Pancaldi2016,Whalen2016}, with $\ge 4$ controlling a typical gene expressed during fly embryogenesis~\cite{Kvon2014}. At the same time, they are selective; thousands have the potential to activate a fly gene encoding an ubiquitously-expressed ribosomal-protein, whilst a different set can act on a developmentally-regulated factor~\cite{Zabidi2015}. In our model, ``red'' enhancers tether ``red'' genic promoters close to ``red'' factories, as ``green'' ones do the same with a different set. (v) Enhancer-target contacts apparently track with the polymerase down the target~\cite{Lee2015}. Thus, when mouse {\it Kit} becomes active, the enhancer first touches the {\it Kit} promoter before contacts move progressively $3'$ at the speed of the pioneering polymerase. This is impossible with conventional models, but simply explained if polymerases transcribing enhancer and target are attached to one factory (Fig. 4Aii,iii). (vi) Single-molecule RNA FISH shows forced looping of the $\beta$-globin enhancer to its target increases transcriptional burst frequency but not burst size~\cite{Bartman2016}, and this general effect is confirmed by live-cell imaging of {\it Drosophila} embryos~\cite{Fukaya2016,Muerdter2016}. Such bursting arises because many ``active'' genes are silent much of the time, and when active they are associated with only one elongating polymerase (Suppl. Note 8). Periods of activity do not occur randomly; rather, short bursts are interspersed by long silent periods. Bursting is usually explained by an equilibrium between ill-defined permissive and restrictive states; we explain it as follows. In Figure 4A, {\it p} often fires when tethered near the factory (giving a burst). Then, once {\it e} terminates, close tethering is lost -- and {\it p} remains silent for as long as it remains far from an appropriate factory. RNA FISH experiments on human {\it SAMD4A} support this explanation; the promoter is usually silent, but adding TNF$\alpha$ induces successive attachments/detachments to/from a factory~\cite{Larkin2013}.

	A related conundrum concerns how super-enhancers (SEs) work. SEs are groups of enhancers that are closely-spaced on the genetic map and often target genes determining cell identity~\cite{Whyte2013,Hnisz2017}. In Figure 4Bi, increasing the number of closely-spaced promoters ({\it e}, {\it e'}, {\it e''}) in the SE increases the time {\it p} spends near a factory (to increase its firing probability).

\subsection{Boundaries}

	TAD boundaries in higher eukaryotes are often marked by CTCF; however, they are also rich in active units marked by polymerase II, nascent RNA, and factors like YY1~\cite{Dixon2012,Rao2014,Weintraub2017}. Similarly, fly boundaries are rich in constitutively-active genes but de-enriched for insulators dCTCF and Su(Hw)~\cite{Ulianov2016,Rowley2017}. Additionally, in yeast (which lacks CTCF), boundaries are often active promoters~\cite{Hsieh2015}. Then, does the act of transcription create a boundary? Studies in {\it Caulobacter crescentus} -- which lacks CTCF but possesses TADs -- shows it does~\cite{Le2016}. For example, in a rich medium, a rDNA gene is a strong boundary; however, this boundary disappears in a poor medium when rRNA synthesis subsides. Inserting active {\it rsaA} in the middle of a TAD also creates a new boundary, and boundary strength progressively falls when the length of the transcribed insert is reduced. We imagine ongoing transcription underlies boundary activity (Fig. 4C).

\section{A great mystery: gene regulation is widely distributed}
 
	Classical studies on bacterial repressors (lambda, lac) inform our thinking on how regulators work: they act locally as binary switches. We assume eukaryotes are more complicated, with more local switches, plus a few global ones (e.g., Oct3/4, Sox2, c-Myc, Klf4). We are encouraged to think this by studies on some diseases~\cite{Deplancke2016}. For example, KLF1 regulates $\beta$ globin expression by binding to its cognate site upstream of the $\beta$-globin gene ({\it HBB}); a C to G substitution at position -87 reduces binding, and this reduces HBB expression and causes $\beta$-thalassaemia. Therefore, we might expect binding of factors to targets drives phenotypic variation. However, results obtained using GWAS (genome-wide association studies) -- an unbiased way of finding which genetic loci affect a phenotype -- lead to a different view for many diseases; they are so unexpected that only general explanations are proffered for them~\cite{Albert2015,Deplancke2016,Boyle2017}.  

\subsection{eQTLs}

	 Quantitative trait loci (QTLs) are sequence variants (usually single-nucleotide changes) occurring naturally in populations that influence phenotypes. Most QTLs affecting disease do not encode transcription factors or global regulators; instead, they map to non-coding regions, especially enhancers~\cite{Javierre2016,Boyle2017}. eQTLs are QTLs affecting transcript levels, and were also expected to encode transcription factors; but again, many do not~\cite{Yvert2003,Boyle2017}. They also map to enhancers~\cite{Boyle2017} and regulate distant genes both cis and trans~\cite{Brynedal2017,GTEx2017,Yao2017}. Additionally, eQTLs and their targets are often in contact~\cite{Javierre2016}, and one trans-eQTL can act on hundreds of genes around the genome -- which often encode functionally-related proteins regulated by similar factors~\cite{Platig2016,Boyle2017,Brynedal2017,Yao2017}. In summary, eukaryotic gene regulation involves distant and distributed eQTLs that look like enhancers. Moreover, copy number of a transcript is a polygenic trait much like susceptibility to type II diabetes or human height -- traits where hundreds of regulatory loci have been identified and where many more await discovery~\cite{GTEx2017}. This complexity is captured by the ``omnigenic'' model, where eQTLs affect levels of target mRNAs indirectly; they modulate levels, locations, and post-translational modifications of unrelated proteins, and these changes percolate throughout the cellular network before feeding back into nuclei to affect transcription of targets~\cite{Boyle2017}. We suggest another -- very direct -- mechanism.

\subsection{A model for direct eQTL action}

	In Figure 5A, all units in the volume determine network structure, and how often each unit visits an appropriate factory; consequently, all units directly affect production of all other transcripts. In other words, gene regulation is widely distributed. A single nucleotide change in enhancer {\it b} (perhaps an eQTL) might reduce binding of a ``yellow'' factor and {\it b}'s firing frequency, and this has consequential effects on how often {\it d} and {\it a} are tethered close to the yellow factory -- and so can initiate. But this change influences the whole network. By altering positions relative to appropriate factories, an eQTL ``communicates'' directly with functionally-related targets, and indirectly (but still at the level of transcription) with all other genes around it in nuclear space. This neatly reconciles how eQTLs target functionally-related genes whilst having omnigenic effects (because targets often share the same specialized factory and nuclear volume, respectively).

\begin{figure*}[!h]
\begin{center}
\includegraphics[width=0.85\columnwidth]{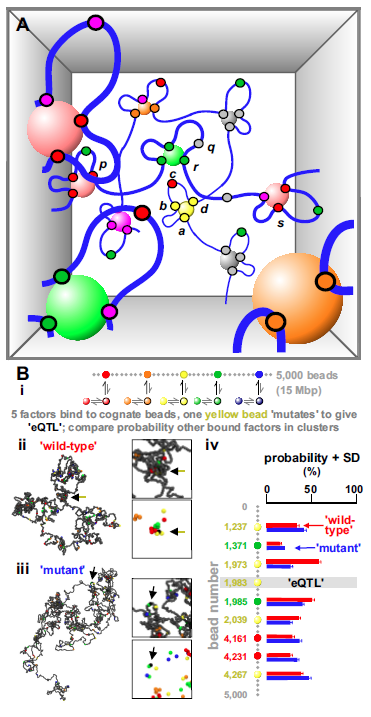}
\end{center}
\caption{
Regulation is widely distributed -- an omnigenic model.
{\bf A.} Activity of every transcription unit (small circles) in the volume depends on the activity of neighbours. {\it b} acts simultaneously as an enhancer of {\it a} and {\it d} (by tethering them close to the yellow factory) and a silencer of {\it c} (by tethering it far from a pink factory). {\it r} acts as a boundary between different TADs containing {\it p} and {\it s}; it also silences {\it q}, by preventing it from accessing a grey factory. Purple units are promiscuous, often initiating in factories of another color.
{\bf B.} Molecular-dynamics simulations of eQTL action. {\bf (i)} Overview. One simulation in a set of $200$ involves $5$ ``factors'' (colored $30$-nm spheres) binding reversibly to cognate beads of similar color randomly distributed along a ``wild-type'' string ($30$-nm bead -- $3$ kbp). Factors can be ``de-phosphorylated/phosphorylated'' to lose/gain affinity at equal rates ($\sim 0.00001$ inverse Brownian times, or $\sim 0.001$ s$^{-1}$). Another set involves a ``mutant'' string with an ``eQTL'' where yellow bead $1983$ becomes non-binding. {\bf (ii,iii)} Snapshots of ``wild-type'' and ``mutant'' fibers (bead $1983$ shown black, arrowed; factors not shown). Boxes: magnifications of regions around bead $1983$ with/without non-binding beads (grey). {\bf (iv)} Positions and colors of all binding beads with altered transcription probabilities. We assume a chromatin bead is transcribed if it is within $54$ nm of a factor of the corresponding color -- when transcribed a bead is also typically in a cluster. Statistical significance for changes in histograms for binding beads shown is calculated assuming Gaussian statistics; histograms are different with $p$-value $p < 0.009$, and $<2$ beads are expected to change this much by chance. 
}
\end{figure*}

	The idea that altering one loop in a network has global effects was tested using simulations of $5$ factors binding to cognate sites in a $5,000$-bead string (Fig. 5Bi; Suppl. Note 6 gives details); as expected, bound factors spontaneously cluster (Fig. 5Bii). We next create an ``eQTL'' in the middle of the (``wild-type'') string by abolishing binding to one yellow bead. This ``mutant'' bead is now rarely in a cluster (Fig. 5Biii, arrow), and it increases or decreases clustering probabilities of many other genes on the string (Fig. 5Biv). 
As clustering determines activity, these simulations provide a physical basis for direct omnigenic effects, and open up the possibility of modeling their action. Results are robust, as, for instance, simulations with different binding affinity, or with factors and binding sites of only a single color, lead to qualitatively similar conclusions.

{{\section{Limitations of the model}}}

        {{Whilst we have seen that the transcription-factory and transcription-factor models can explain many disparate observations, from phase separation of active and inactive chromatin through to eQTL action, this review would not be complete without a critical discussion of their limitations. Besides the complicated relation between TADs and transcription already reviewed, we list here some other challenges to our model.}}

        {{First, the simplest version of our model does not immediately account for the bias in favor of convergent CTCF loops (over divergent ones) -- which is naturally explained by the ``loop-extrusion'' model~\cite{Nasmyth2011,Sanborn2015,Fudenberg2016,Brackley2017b} (see also Suppl. Note 2). However, the loop-extrusion and transcription-factor model are not alternative to one another, but complementary, so convergent loops are naturally recovered by a combined model where chromosomes are organized by both transcription factors and cohesin (bioRxiv: https://doi.org/10.1101/305359). Additionally, the motor activity behind loop extrusion, if present, may be provided by transcription itself~\cite{Racko2017} (Suppl. Note 2).}}

        {{Second, the structures of mitotic and sperm chromatin pose a challenge to all models (Suppl. Notes 9 and 10). For ours, it is difficult to reconcile the persistence of loops during these stages with the common assumption that all factors are lost from chromatin. However, recent results suggest this assumption is incorrect, and that many factors do actually remain bound in mitosis~\cite{Teves2016} (Suppl. Note 9).
The case of sperm is harder to explain. We speculate cohesin and other factors may still operate, and this might be sufficient to explain the observations (Suppl. Note 10).}}\\

\section{Conclusion}

	Seeing is believing. While clusters of RNA polymerase II tagged with GFP are seen in images of living cells~\cite{Sugaya2000,Cisse2013,Chen2016,Cho2016a,Cho2016b}, decisive experiments confirming ideas presented here will probably involve high-resolution temporal and spatial imaging of single polymerases active on specified templates. But these are demanding experiments because it is so difficult to know which kinetic population is being imaged. For example, an inactive pool of polymerase constitutes a high background; $\sim 80\%$ is in a rapidly-exchanging pool, and so soluble or bound non-specifically~\cite{Kimura1999}. If mammalian polymerases are like bacterial ones, most at promoters fails to initiate, and -- of ones that do initiate -- $99\%$ abort within $\sim 10$ nucleotides to yield transcripts too short to be seen by RNA-seq~\cite{Goldman2009}. Then, eukaryotic enzymes on both strands abort within $20-500$ nucleotides to give products seen by RNA-seq as promoter-proximal peaks~\cite{Ehrensberger2013}. On top of this, $\sim 60\%$ further into genes pause for unknown periods~\cite{Day2016}. We may also think that active and inactive polymerases are easily distinguished using inhibitors, but DRB and flavopiridol do not block some polymerases at promoters (e.g., ones phosphorylated at Ser5 of the C-terminal domain), $\alpha$-amanitin takes hours to act, and both $\alpha$-amanitin and triptolide trigger polymerase destruction~\cite{Bensaude2011}.

        In biology, structure and function are inter-related. Here, we suggest that many individual acts of transcription determine global genome conformation, and this -- in turn -- feeds back to directly influence the firing of each individual transcription unit. Consequently, ``omnigenic'' effects work both ways. [Note the term ``omnigenic'' is used here to include both genic and non-genic transcription units.] In other words, transcription is the most ancient and basic driver of the organization in all kingdoms, with recently-evolved factors like CTCF modulating this basic structure. It also seems likely that transcription factories nucleate related ones involved in replication, repair, and recombination~\cite{Papantonis2013}, as well as organizing mitotic chromosomes (Suppl. Note 9). They may also play important roles in other mysterious processes like meiotic chromosome pairing and transvection~\cite{Xu2008b}.

\section{ACKNOWLEDGEMENTS}

This work was supported by the European Research Council (CoG 648050, THREEDCELLPHYSICS; DM), and the Medical Research Council (MR/KO10867/1; PRC). We thank Robert Beagrie, Chris A. Brackley, Davide Michieletto and Akis Papantonis for helpful discussions. 

\subsubsection{Conflict of interest statement.} None declared.

\clearpage

\newpage


\section*{Supplementary Notes}

\subsection*{Supplementary Note 1: Some properties of loops known before the invention of 3C}

The idea that chromatin fibers are looped is an old one. Extended lampbrush loops were first described by Flemming in the 1880's~\cite{Gall1996,Morgan2002}. Flemming carefully spread what we now call chromosomes of amphibian oocytes (at the stage when parental homologs pair during meiosis), and saw that most chromatin was visibly looped. In the 1970's, the genome of {\it Escherichia coli} -- which had a circular genetic map -- was also shown to be looped. Bacteria were lysed in a high salt concentration that stripped off proteins to leave naked DNA still associated with a cluster of engaged RNA polymerases~\cite{Stonington1971}; this DNA was supercoiled -- and so looped (as supercoils are lost spontaneously from linear fibers~\cite{Worcel1972}). Then, analogous experiments on human cells gave the same result; this indicated that even DNA of organisms with linear genetic maps was looped~\cite{Cook1975}. Moreover, looping and transcription were tightly correlated, as supercoils progressively disappear when transcriptionally-active chicken erythroblasts mature into inactive erythrocytes~\cite{Cook1976}. Additional evidence for looping came from analyses of rates at which nucleases and $\gamma$-rays cut fibers; supercoils are released by one cut, but two nearby cuts are required to release DNA fragments from nuclei~\cite{Cook1976,Igo1978}. 

	Loops seen in these biochemical studies might have been generated artifactually during lysis. This provoked development of gentler methods that used ``physiological'' buffers and conditions where polymerases ``ran-on'' at rates found in vivo; then, it was likely that structure is preserved if function is also preserved. Loops under such conditions were characterized in detail, and by 1990 ($>10$ y before the invention of 3C) it was known that essentially all chromatin in active nuclei of men, mice, flies, and yeast was looped, and that promoters and active transcription units were major anchors (reviewed in~\cite{Papantonis2013}). In interphase HeLa cells, the average contour length is $\sim 86$ kbp, with this average covering a wide range from $12.5-250$ kbp~\cite{Jackson1990}.

	As discussed in the main text, improvements in Hi-C resolution allow detection of loops anchored by convergent CTCF sites~\cite{Rao2014}. However, many of these loops are longer than the longest described above. Moreover, the early biochemical studies showed that loops persist during mitosis (see~\cite{Jackson1990} and Supplementary Note 9); this contrasts with the failure of Hi-C to detect loops at this stage (presumably tight packing creates additional contacts that obscure ones due to looping). While Hi-C remains a powerful tool for detecting loops, it seems we must await further improvements in resolution before it is able to detect many loops in many organisms.



\subsection*{Supplementary Note 2: The ``loop-extrusion'' model, and other mechanisms driving enlargement of contact loops stabilized by CTCF/cohesin}

	Various mechanisms could enlarge contact loops once binding of the cohesin ring generates a small loop. We begin by noting that it remains uncertain whether cohesin stabilizes loops by acting as one ring embracing two fibers, or two connected rings each embracing one (Supplementary Fig. S2Bi; see~\cite{Nasmyth2011}). Whatever the structure, a small loop can only enlarge if the cohesin ring (or rings) translocate down the fiber(s). This can be achieved in various ways. First, cohesin could possess an inbuilt motor (Fig. 1C); {{this assumption underlies the ``loop-extrusion model''~\cite{Alipour2012,Sanborn2015,Fudenberg2016}}}.  {{This assumption is based on the fact that cohesin}} is an ATPase~\cite{Nasmyth2011}, and that some of its relatives are known motors~\cite{Eeftens2017,Terakawa2017,Ganji2018}. For example, SMC (structural maintenance of chromosomes) complexes may travel at $\sim 50$ kbp/min in living bacteria~\cite{Wang2017a}, and yeast condensin moves $\ge 10$ kbp mainly in one direction at $\sim 4$ kbp/min~\cite{Terakawa2017}. However, if a motor, cohesin would have to be more processive and faster than RNA polymerase to extrude a $1$-Mbp loop in $\sim 25$ min (its average residence time on DNA). Second, a motor like RNA polymerase could push cohesin along a fiber directly~\cite{Busslinger2017}, or generate the supercoils that do so indirectly~\cite{Racko2017}. Third, diffusion could underlie the motion (Supplementary Fig. S2B,ii; see~\cite{Brackley2017b}). At first glance, this seems an oxymoron -- 1D diffusion gives a bi-directional random walk and not the uni-directional motion required for extrusion. However, a random walk can be biased by loading a second ring to limit movement of the first back towards the loading site; then, the second ring exerts an effective osmotic pressure that rectifies diffusion of the first. Simulations confirm this, and show that loading more rings leads to their clustering behind the pioneer. Then, if one ring in a cluster dissociates, the remainder can maintain extrusion until bound CTCF stalls it. Such molecular ratchets provide viable mechanisms driving extrusion in the required time -- without invoking motors. Additionally, loop formation need not arise from unidirectional extrusion: if cohesin sticks strongly to CTCF once it finds it by diffusive sliding; this is enough to explain the formation of convergent loops~\cite{Brackley2017b}.

        {{As shown in Figure~2C in the main text, loop extrusion through cohesin rings in mammals seems to stall at CTCF bound to convergent cognate sites, and we would expect this to be so whether or not the CTCF is in a transcription factory. Consequently, loop extrusion and its stalling at such sites may in principle be readily accommodated within our model.}}

\subsection*{Supplementary Note 3: Most transcription occurs in factories}

	Some cars are assembled by enthusiasts at their own homes, but most are made in factories; are most transcripts made in factories? The answer came after permeabilizing HeLa cells in a physiological buffer (see Supplementary Note 1), labeling nascent RNA by ``running-on'' in biotin-CTP or Br-UTP, and immuno-labeling the resulting biotin- or Br-RNA~\cite{Jackson1993a,Papantonis2013}. Here, the challenge is to ensure that signals seen inside and outside factories accurately reflect relative amounts of transcription occurring in the two places. How can one ensure this? The answer is to run-on for longer under conditions where signal in factories grows stronger without more factories being detected (which indicates all factories are being seen), as extra-factory signal remains at background levels (indicating this signal is not due to incorporation of labelled precursors by polymerases outside factories). Quantitative light and electron microscopy (often using thin 100 nm sections to improve z-axis resolution) showed that at least $92\%$ signal was in factories~\cite{Pombo1999,Faro2006}. As experiments involving different labels, antibodies, and detection systems gave similar results, it seems that essentially all transcription occurs in factories.

\subsection*{Supplementary Note 4: Some characteristics of factories in HeLa and HUVECs}

	Factories in sub-tetraploid HeLa and diploid HUVECs are the best characterized~\cite{Papantonis2013}. A typical nucleolar factory in HeLa (i.e., a fibrillar center or FC, plus $4$ associated dense fibrillar components or DFCs) contains $\sim 4$ rDNA templates each packed with $\sim 125$ active molecules of RNA polymerase I. We imagine a promoter snakes over the surface of the FC -- a cluster of polymerase I and its upstream binding transcription factor, UBF. After the promoter initiates, the polymerase extrudes the promoter -- which re-initiates when it reaches the next polymerase on the surface. Extruded transcripts then form the DFC. Stripping off template and transcript from the surface gives the ``Christmas tree'' seen in spreads (Fig. 3Ci). Finally, transcripts from one or more FCs and DFCs are assembled into ribosomes in the surrounding granular component.
 
	The general structure of nucleoplasmic factories is like that of nucleolar ones, with nascent transcripts again found on the surface of a central core~\cite{Papantonis2013}; now however, most active genes are productively transcribed by only one active polymerase and not the many seen on active ribosomal cistrons (see Supplementary Note 8). Thus, in a dividing HeLa cell, nascent nucleoplasmic RNA is found on the surface of a protein-rich factory core (diameter $50-175$ nm; mass $\sim 10$ MDa). This core has a mass density $\sim 0.1\times$ that of a nucleosome, and so is likely to be porous. There are $\sim 6,000$ polymerase II factories per nucleus (density $\sim 9.3$ factories/$\mu$m$^3$; inter-factory spacing $\sim 220-475$ nm), with each factory containing $\sim 10$ active polymerases (the remaining $\sim 80\%$ of nuclear polymerase constitutes the inactive and rapidly-exchanging soluble pool). There are also $\sim 1,200$ polymerase III factories with slightly smaller diameters. These different factories have been partially purified and their proteomes and transcriptomes analyzed; they contain the expected polymerases, associated factors, and nascent RNAs~\cite{Melnik2011,Caudron2015}. 

	In a starved HUVEC in G0 phase (which has a smaller nucleus than HeLa), there are $\sim 2,200$ polymerase II factories, and so $\sim 30$ in a territory occupied by a $100$-Mbp chromosome. After treatment with TNF$\alpha$ (tumor necrosis factor $\alpha$) for $30$ min, there are a hundred or so specialized ``NF$\kappa$B'' factories per nucleus (but not more than $\sim 250$~\cite{Papantonis2012}). These numbers mean a typical gene responding to the cytokine has a good chance of visiting several ``NF$\kappa$B'' factories every few minutes by diffusion.

\subsection*{Supplementary Note 5: Some evidence supporting the idea that active polymerases do not track}

	The extensive evidence that active polymerase do not track has been reviewed~\cite{Papantonis2013}; three kinds are briefly summarized here. First, if active RNA polymerases track, exhaustive treatment with endonucleases should detach most DNA in a loop from tethering points; consequently, three markers of the active complex -- the tracking polymerase, transcribed template, and nascent RNA -- should all be detached from tethering points (Supplementary Fig. S4Ai). This experiment gave unexpected results: transcribed templates and nascent RNAs were not detached, and this pointed to active polymerases being at tethering points and so probably immobilized there (Supplementary Fig. S4Aii; see~\cite{Jackson1981}). But perhaps active enzymes precipitate on to the underlying nuclear sub-structure in the unphysiological buffer used, to form new (artefactual) anchors that did not exist previously? However, using the ``gentle'' conditions described in Supplementary Note 1 gave the same result; removing the body of loops still did not remove any of the three markers. Instead, all remained. This again implied that active polymerizing complexes are significant tethers~\cite{Jackson1985,Dickinson1990}, and fine-structure mapping confirmed this~\cite{Jackson1993b}. 

	The second kind of evidence involved analysis of 3C contacts made between one short gene and one very long gene -- $11$-kbp {\it TNFAIP2} and $221$-kbp {\it SAMD4A}; both genes respond to TNF$\alpha$, and the short one is used as a reference point~\cite{Papantonis2010}. Before adding TNF$\alpha$, both are transcriptionally silent and rarely contact each other (both roam ``outer space''; Supplementary Fig. S4B, 0 min). After adding TNF$\alpha$, contacts change in a way impossible to reconcile with a model involving tracking polymerases (Supplementary Fig. S4B, $10-85$ min). Thus, within $10$ min, the reference point (i.e., {\it TNFAIP2}) often contacts the {\it SAMD4A} promoter. After $30$ min, it no longer contacts the {\it SAMD4A} promoter; instead, it contacts a point one-third of the way into the long gene. After $60$ min, contacts shift two-thirds into {\it SAMD4A}, and after $85$ min they reach the terminus. Such results are simply explained if polymerases active on the two genes are immobilized in one ``NF$\kappa$B'' factory. After $10$ min, both genes attach to (and initiate in) such a factory; consequently, promoter—promoter contacts are seen. As {\it SAMD4A} is so long, the polymerase takes $85$ min before it reaches the terminus. In contrast, a polymerase on {\it TNFAIP2} terminates within minutes, and the short gene then goes through successive transcription cycles -- sometimes attaching to (and detaching from) the same factory. If it reinitiates after $30$, $60$, or $85$ min in the same factory (when the pioneering polymerase on {\it SAMD4A} has transcribed one-third, two-thirds, or all of the way along the long gene), it will contact points on {\it SAMD4A} that become progressively closer to the terminus -- as is seen. RNA FISH coupled to super-resolution localization confirms this interpretation: intronic (nascent) RNAs copied from relevant segments of the two genes lie close enough together at appropriate times to be on the surface of one spherical factory with a diameter of $\sim 90$ nm. Immobilization of polymerases also provides a simple explanation for the way {\it e}-{\it p} contacts apparently track downstream of {\it p} with the polymerase in Figure 4A (panels ii, iii).

	The third kind of evidence involves real-time imaging of the human gene encoding cyclin D1 and its transcript as the gene becomes active~\cite{Germier2017}. Thus, addition of estrogen switches on transcription in minutes, and this correlates with a reduction in the volume explored by the gene. Inhibitor studies show the constrained mobility depends on transcriptional initiation. This confirms that genes become highly confined when active.

	Evidence often cited in favor of tracking polymerases comes from images of lampbrush loops. Like ``Christmas trees'' in ``Miller'' spreads (Fig. 3Ci), lampbrush loops are made by spreading a 3D structure; active polymerases and nascent RNAs (detected by immuno-labeling and autoradiography, respectively) are seen out in loops in 2D spreads~\cite{Morgan2002,Gall2016}. However, transcription is required to form and maintain loops seen after spreading~\cite{Gall1998}. In addition, both markers are even more concentrated in the axial chromomeres to which loops are attached~\cite{Gall1998,Snow1969}, and no loops are seen in whole-cell sections where chromatin appears as a granular aggregate~\cite{Mott1975}. As with ``Christmas trees'', we suggest active polymerases are stripped off factories during spreading; significantly, possible intermediates in such a process -- large granular aggregates -- are often seen attached to spread loops~\cite{Mott1975}. Consequently, these images do not provide decisive evidence for the traditional model. 

\subsection*{Supplementary Note 6: Details of simulations}

	Results in Figures 5B and Supplementary Figure S6B were obtained using Brownian dynamics (BD) simulations. These were run with the LAMMPS (Large-scale Atomic/Molecular Massively Parallel Simulator) code~\cite{Plimpton1995}, by performing molecular dynamics simulations with a stochastic thermostat~\cite{Brackley2016}. Chromatin fibers are modeled as bead-and-spring polymers using FENE bonds (maximum extension $1.6$ times bead diameter) and a bending potential that allows persistence length to be set (here $3$ times chromatin-bead size, corresponding to a flexible polymer). Protein--protein and template--template interactions involve only steric repulsion. For template--protein interactions, we used a truncated and shifted Lennard-Jones potential (detailed below). All participants are confined within a cube with periodic boundary conditions, but strings are ``unwrapped'' for presentational purposes (i.e., disconnected strings are rejoined). In all cases, simulations are initialized with chromatin fibers as random walks and proteins distributed randomly with uniform density over the simulation domain. Any overlap between beads (proteins or chromatin) are eliminated with a short equilibration run with soft repulsive interactions between any two beads. Length and time scales in simulations can be mapped to physical ones, for example, by identifying bead size as $30$ nm (representing $3$ kbp), and a time simulation unit as $0.01$ s (this unit corresponds to the square of the bead size over the diffusion coefficient of a bead in isolation; see~\cite{Brackley2016,Michieletto2017}).

	For Figure 5B, we consider $5$ different factors (red, green, blue, orange and yellow) that can bind specifically to $5$ sets of cognate sites (of the same color) scattered randomly along a chromatin fiber of $5,000$ beads. The fiber represents $15$ Mbp, and colored beads (cognate binding sites for factors) are spaced -- on average -- every $30$ beads (colored beads are assigned a random color between red, green, blue, orange and yellow, with equal probability). In the set of simulations presented in Figure 5B, there are in total $172$ coloured chromatin beads, of which $39$ are red, $38$ green, $32$ blue, $33$ orange and $30$ yellow. The $5$ factors also bind non-specifically to every other (non-colored) bead. Specific (non-specific) interaction between chromatin and protein are modeled as truncated-and-shifted Lennard-Jones potentials with interaction energy $7.1$ ($2.7$) $k_BT$, with an interaction range of $54$ nm. We assume factors switch between binding and non-binding states at rate $\alpha = 10^{-3}$~\cite{Brackley2017a}.  Data presented in the histogram were averaged over $200$ simulations, each of $10^5$ time units. In snapshots shown, only the fiber (and only the $5$ sets of cognate sites) are shown for clarity.

	For Supplementary Figure S6B, we consider a single type of (non-switching) factor (so $\alpha = 0$), binding only specifically to regularly-spaced cognate sites (modeled as for Figure 5B).

        For both cases, additional simulations with different interaction energy and range for DNA-protein interactions show the results to be qualitatively robust, provided that the interaction leads to multivalent binding. For Figure 5B, we have also run additional simulations with factors and binding sites of a single color, and found similar results when simulating eQTL action. Additionally, simulations with similar number of factors, but no switching give again qualitatively similar results -- in this case, the protein clusters are much less dynamic as expected. 

\subsection*{Supplementary Note 7: Some additional conundrums -- transcriptional interference, clustering of co-regulated genes, assembly of nuclear bodies}

	In the phenomena of ``transcriptional interference'', firing of one promoter prevents firing of an adjacent one; this has been difficult to explain because interference extends over at least $10$ kbp~\cite{Emerman1986}. The model and data illustrated in Figure 3D provide a simple explanation for the phenomenon. Thus, when promoter {\it p} is positioned anywhere in the black part of the fiber (Fig. 3Di), the fiber cannot bend back to allow {\it p} to reach the green volume on the surface of the factory; consequently, transcription of {\it e} ``interferes'' with (i.e., prevents) {\it p} from firing whilst {\it e} remains tethered to the factory.

	In bacteria, co-regulated operons lying $>100$ operons apart on the genetic map nevertheless often contact each other in 3D space~\cite{Xie2015}. In man, co-functional genes are also concentrated on the genetic map and in nuclear space~\cite{Thevenin2014}. What underlies this clustering, for which there seems to be no explanation? We suggest evolutionary pressures broadly concentrate co-regulated genes on the genetic map so they can easily access appropriate factories {{(Supplementary Fig. S5C)}}.

	How might functional nuclear bodies form? The nucleolus is both the prototypic factory and nuclear body. Nucleoli spontaneously assemble in human fibroblasts around tandem repeats inserted ectopically if repeats encode binding sites for UBF (the major transcription factor used by polymerase I); resulting ``pseudo-nucleoli'' contain UBF. If inserts also encode rDNA promoters, resulting ``neo-nucleoli'' contain active polymerase~\cite{Grob2014a}. Histone-locus bodies (HLBs) in {\it Drosophila} illustrate assembly of polymerase II factories. Replication-coupled histone genes are encoded by $\sim 100$ $5$-kbp repeats, each with $5$ histone genes, with transcription of H3 and H4 being driven by one bidirectional promoter. Ectopic insertion of $297$ bp from this promoter leads to HLB assembly~\cite{Salzler2013}. We again suggest that the act of transcription underlies the clustering of polymerases/factors into specialized factories and the assembly of nuclear bodies -- via the bridging-induced attraction (i.e., the process illustrated in Fig.~2).

\if{
\subsection*{Supplementary Note 9: The relationship between TADs and transcription}

	Various studies address this issue, and give conflicting results. For example, in mouse neural progenitor cells, one of the two X chromosomes is moderately compacted and largely inactive. Inactive regions do not assemble into A/B compartments or TADs, unlike active ones. Moreover, in different clones, different regions in the inactive X escape inactivation, and these form TADs~\cite{Giorgetti2016}. Here, structure and activity are tightly correlated (in accord with our model). Similarly, inhibiting transcription in the fly leads to a general reorganization of TAD structure, and a weakening of border strength~\cite{Li2015}. However, another study points to some TADs appearing even though transcription is inhibited~\cite{Hug2017}. After fertilization, the zygotic nucleus in the fly egg is transcriptionally inactive. As the embryo divides, zygotic genome activation occurs so that by nuclear cycle $8$ (nc8), $\sim 180$ genes are active, and these seem to nucleate a few TADs detected at nc12 (so transcriptional onset and TAD appearance correlate -- again in accord with our model). As more genes become active at nc13, $3$-fold more TADs develop by nc14, and polymerase II plus Zelda (a zinc-finger transcription factor) are at boundaries (again a positive correlation). If transcriptional inhibitors are injected into embryos before nc8, boundaries and TADs seen at nc14 are less prominent, but some TADs still develop (implying TADs appear independently of transcription). However, interpretation is complicated. Although inhibitors reduce levels of $5$ mRNAs already being expressed, they only slightly affect levels of polymerase II bound at the $5'$ end of genes expressed at nc14; this indicates that inhibition is inefficient, so it remains possible that the remaining transcription stabilizes the TADs seen.
 
	Studies on mouse eggs and embryos also provide conflicting data. Thus, activity is lost as oocytes mature, and TADs plus A/B compartments disappear~\cite{Du2017,Flyamer2017,Ke2017}; therefore, loss of structure and activity again correlate (consistent with our model). After fertilization, the zygote contains two nuclei with different conformations; both contain TADs, but the maternal one lacks A/B compartments. Then, as transcription begins, TADs appear (again a positive correlation), but $\alpha$-amanitin (a transcriptional inhibitor) does not prevent this~\cite{Du2017,Ke2017}. However, interpretation is complicated: $\alpha$-amanitin acts notoriously slowly~\cite{Bensaude2011}, and inhibition was demonstrated indirectly (levels of steady-state poly(A)$^{+}$ RNA fall, but reduction of intronic RNA would be a more direct indicator of inhibition).
 
	These results show that the appearance of TADs generally correlates with transcriptional activation, but the correlation breaks down when transcriptional inhibitors are used (with the caveat that inhibitors might be ineffective). In our model, TADs reflect the underlying network of loops stabilized by factors as well as polymerases, so some loops and clusters will form (giving TADs) even when polymerases are inhibited. Our model also seems to conflict with claims that TADs are invariant between cell types~\cite{Dixon2012,Nora2012}; nevertheless, there are always slight differences that could reflect slight differences in expression profiles.}
\fi

\subsection*{Supplementary Note 8: Most active genes are associated with one productively-elongating polymerase}

	Many studies indicate so-called ``active'' genes are silent much of the time, and when active they are associated with only one productively-elongating polymerase -- even in bacteria (reviewed in~\cite{Finan2011}). For example, a comprehensive survey of RNA synthesis and degradation in mouse fibroblasts shows $\sim 2$ mRNAs are produced per ``active'' gene per hour (range $\sim 0.2-20$~\cite{Schwanhausser2011}). As polymerase II copies at $\sim 3$ kbp/min and a typical gene is $\sim 30$ kbp, copying occurs for only $\sim 20$ min in every hour -- or one-third of the time. Of course, longer genes have a greater chance of being associated with $>1$ polymerase~\cite{Larkin2012,Larkin2013}, and one rRNA gene can be transcribed simultaneously by $>100$ molecules of a different polymerase -- RNA polymerase I (Fig. 3C).

\subsection*{Supplementary Note 9: The persistence of loops during mitosis}

	How interphase structures change during mitosis is one of the oldest challenges in biology, and remains one today. For example, early biochemical studies showed that loops in interphase HeLa persist into mitosis without change in contour length (Supplementary Note 1; see~\cite{Jackson1990}). However, no loops, TADs, or A/B compartments are seen by Hi-C in mitotic human cells~\cite{Naumova2013}. That loops are missed is unsurprising: resolution is insufficient against the high background induced by close packing. That A/B compartments go undetected is surprising, as Giemsa bands seen in karyotypes are such close structural counterparts (presumably they are missed because resolution is again insufficient).
 
	The persistence of loops presents a challenge to all models -- and particularly ours -- as it is widely assumed that the players stabilizing loops (which might be CTCF in some models, or polymerases/factors in ours) dissociate during mitosis. Consequently, loops should disappear (as indicated by Hi-C data), or other players must take over to stabilize them (if so, what are these players?). However, recent findings suggest the underlying assumption is incorrect. Thus, many genes turn out to be transcribed during mitosis, albeit at lower levels~\cite{Palozola2017}, so some polymerases and factors must remain bound. Moreover, some genes and enhancers even become more active, and global levels of active marks (e.g., H3K4me2, H3K27ac) also increase~\cite{Liang2015,Liu2017}. Significantly, live-cell imaging shows that many GFP- and halo-tagged factors (e.g., Sox2, Oct4, Klf4, Foxo1/3a) -- including ones previous immunofluorescence studies had shown to be lost -- actually remain bound. The (apparent) loss was traced to a fixation artifact; as the fixative (paraformaldehyde) enters cells, it removes factors from the soluble pool to bias exchange with bound ones, and this strips bound molecules from chromosomes~\cite{Teves2016}. Since we now know polymerases and factors do persist, they can remain the structural organizers during mitosis. In addition, they can also ``bookmark'' previously-active genes for future activity when chromosomes re-enter interphase~\cite{Salzler2013,Grob2014a,Grob2014b,Hsiung2016}.

\subsection*{Supplementary Note 10: The structure of transcriptionally-inert sperm chromatin}

	{{The transcriptionally-inactive sperm nucleus has traditionally been viewed as a mass of unstructured and highly-compacted fibers of protamine and DNA. However, recent work on mammalian sperm shows these fibers to be far from featureless at both local and global levels. For example, their (poised) promoters and enhancers carry active marks and positioned nucleosomes reminiscent of those found in their precursors (i.e., round spermatids) and ES cells, and Hi-C analysis yields A/B compartments and TADs often defined by bound CTCF~\cite{Battulin2015,Jung2017}. These findings represent a challenge for all models, and we now offer some speculations on how they might be accommodated by ours. Thus, we assume that during development of sperm, polymerases become inactive as protamines collapse pre-existing loops around factories; then, local marks, TADs, and A/B compartments would persist. Alternatively (or additionally), some polymerases might remain active as they do in mitosis (Supplememental Note 9).}}

\bibliography{references}

\newpage


\vspace{\fill}

\onecolumngrid

\begin{minipage}{\linewidth}
\centering
\includegraphics[width=0.5\textwidth]{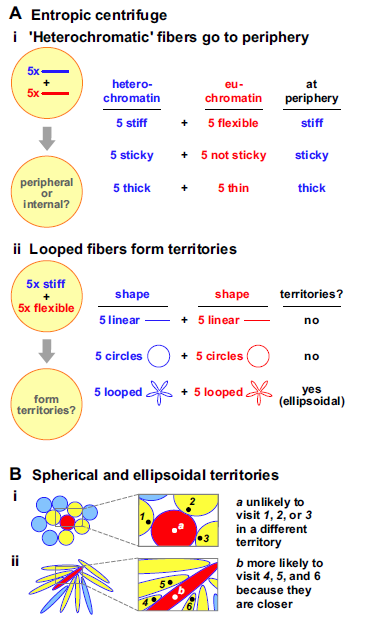}
\captionof{figure}{\small An entropic centrifuge positions and shapes chromatin fibers.
{\bf A.} Monte Carlo simulations involved two sets of $5$ fibers ``diffusing'' in a sphere, and determination of ultimate positions and shapes~\cite{Cook2009}.
{\bf (i)} ``Heterochromatic'' v ``euchromatic'' sets; heterochromatic fibers with higher stiffness, stickiness for others of the same type, and thickness tend to end up at the periphery. 
{\bf (ii)} Stiff v flexible sets (linear, circular, or looped); only looped fibers form territories (others intermingle).
{\bf B.} Ten ellipsoids (``territories'') pack together more tightly than $10$ spheres of similar volume, and may contact more neighbors; they are also less likely to become locally jammed because they have one thinner axis and so can escape through smaller gaps~\cite{Donev2004,Man2005}. For example, consider an ellipsoidal territory (principle axes $1:2.9:4.5$) and a spherical one of similar volume (diameter $4$ $\mu$m). Then, $22\%$ of the ellipsoidal volume is within $125$ nm of the surface compared to $18\%$ of the spherical one, and the average shortest path of any point in the ellipsoid to the surface is $300$ nm (i.e., $60\%$ of the shortest path in the sphere; calculated as described in~\cite{Hart1994}). Ellipsoidal territories are found in haploid mouse embryonic stem (ES) cells~\cite{Stevens2017}, NIH 3T3 cells (principle axes $1:2:3.5$ or $1:1.6:2.3$ depending on substrate~\cite{Wang2017b}), and pro-B nuclei (principle axes $1:2.9:4.5$~\cite{Khalil2007}). 
{\bf (i)} The red sphere touches $4$ yellow ones.
{\bf (ii)} The red ellipsoid touches $7$ yellow ones, and {\it b} at its center is closer to {\it 4}, {\it 5}, and {\it 6} than {\it a} is to {\it 1}, {\it 2}, and {\it 3}. }
\end{minipage}

\vspace{\fill}

\begin{minipage}{\linewidth}
\centering
\includegraphics[width=0.5\textwidth]{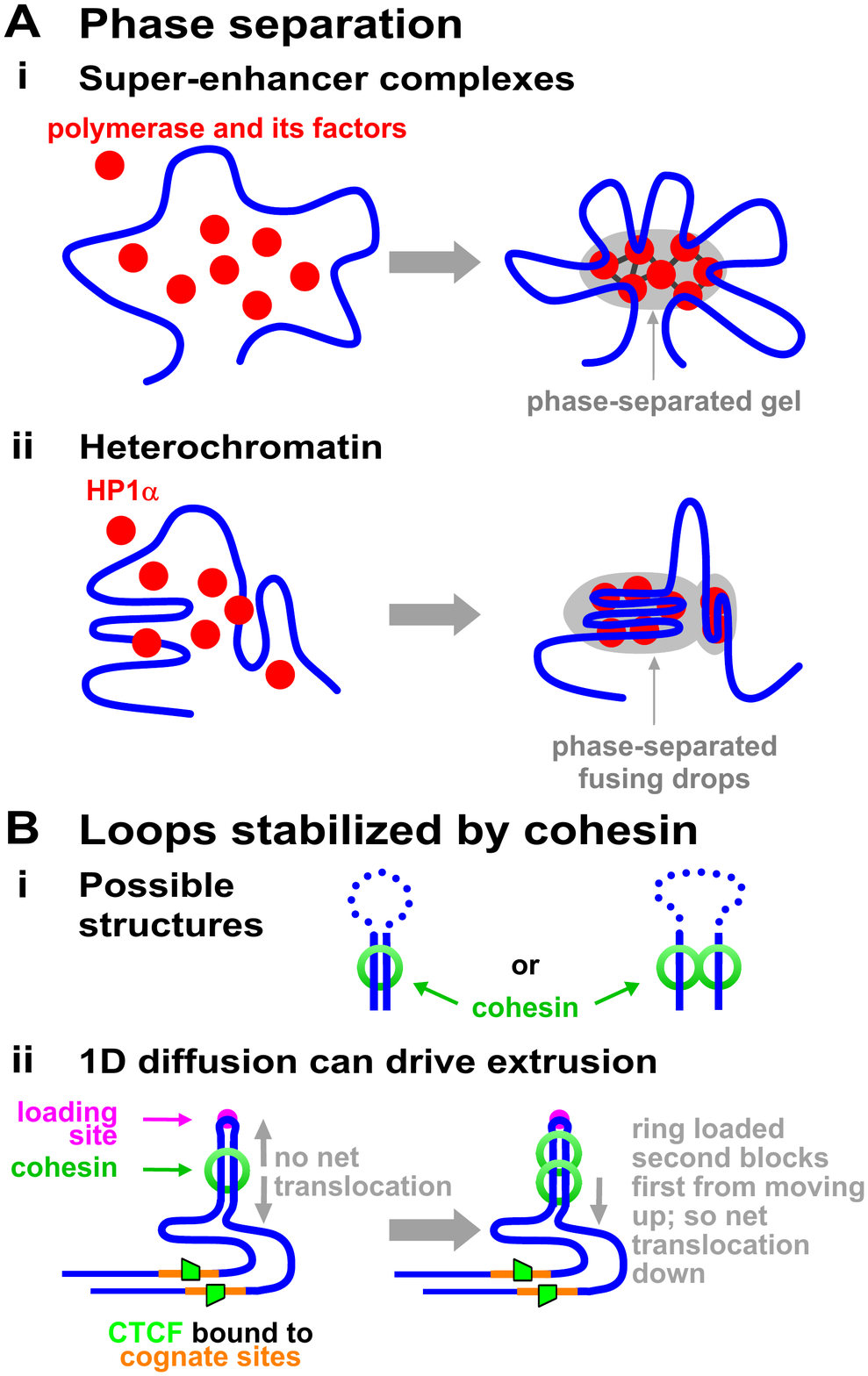}
\captionof{figure}{\small 
Some mechanisms creating loops.
{\bf A.} Phase separation. 
{\bf (i)} Super-enhancer complexes~\cite{Hnisz2017}. The polymerase and its factors bind to promoters and form a phase-separated cluster or gel stabilized by multivalent interactions (black lines); this cluster/gel organizes surrounding loops. This structure is essentially the same as that of a transcription factory. 
{\bf (ii)} Heterochromatin. HP1$\alpha$ forms (phase-separated) liquid-like drops if local concentrations are high enough; it staples fibers together into compact structures with mini-loops~\cite{Larson2017,Strom2017}. Here, two liquid drops have just fused to compact two heterochromatic regions.
{\bf B.} Stabilizing loops with cohesin, and enlarging them by 1D diffusion. 
{\bf (i)} Two possible arrangements for a loop stabilized by cohesin; we assume here that one cohesin ring embraces two duplexes (left), but the same argument applies if two rings each embrace one duplex (right).
{\bf (ii)} A loop stabilized by cohesin could enlarge by 1D diffusion as follows. After binding to the loading site, cohesin then diffuses in a 1D random walk along the fiber; consequently, there is no net translocation along the fiber, and the loop does not enlarge. However, this random walk is biased if a second ring loads at the same loading site, as the second now limits movement of the first back towards the loading site. In practice, the second exerts an effective osmotic pressure that rectifies diffusion of the first. This molecular ratchet provides a viable mechanism driving extrusion without the need to invoke a motor.
}
\end{minipage}

\vspace{\fill}

\begin{minipage}{\linewidth}
\centering
\includegraphics[width=\textwidth]{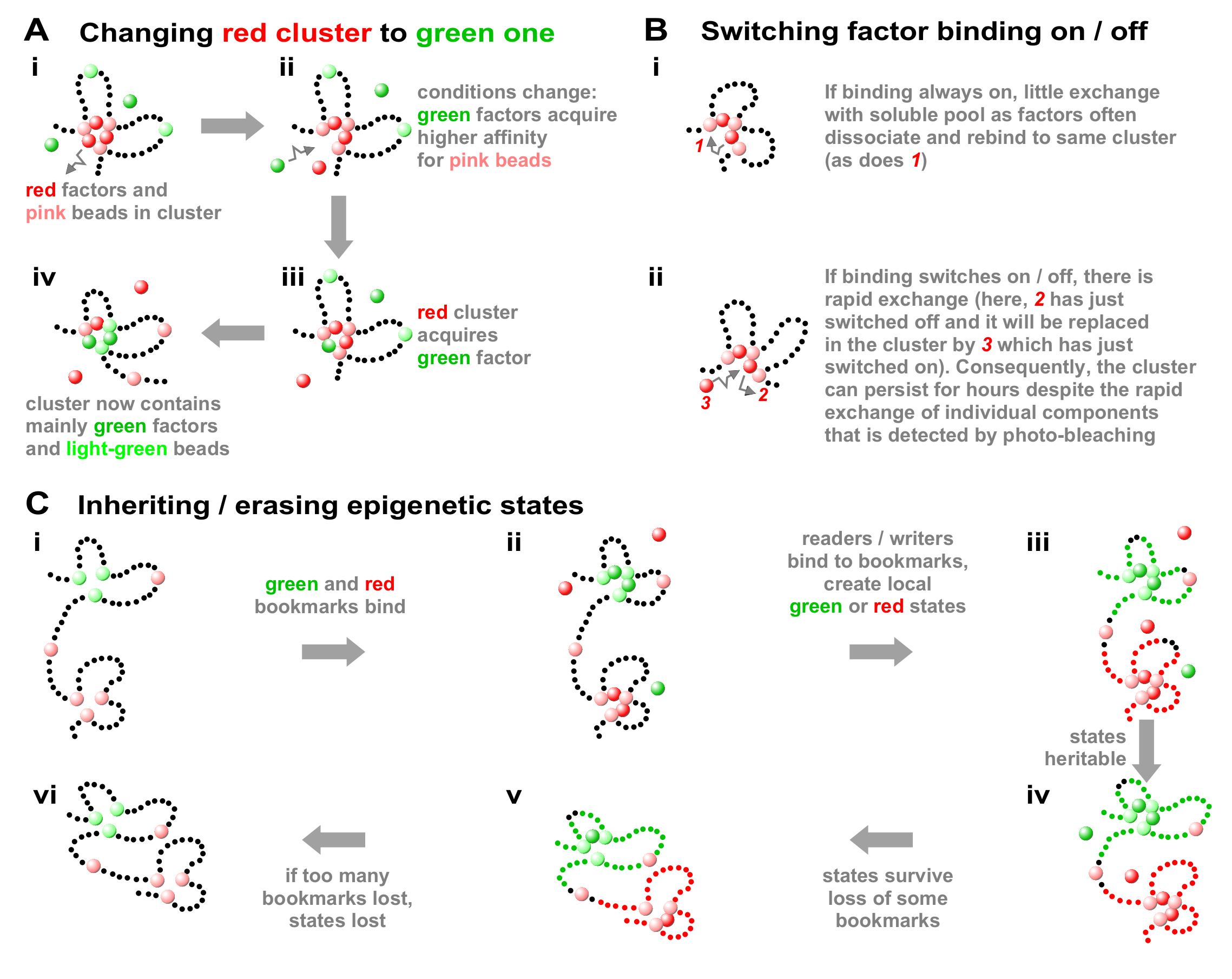}
\captionof{figure}{\small
Cluster growth and stability seen in Brownian-dynamics simulations of chromatin.
{\bf A.} Cluster ``differentiation''; pink/light-green beads represent genes expressed before/after differentiation~\cite{Brackley2016}.
{\bf (i)} Different factors (green, red spheres) bind to cognate sites (light-green, pink beads). Initially, green factors have no affinity for any bead, but red factors can bind to pink ones; red clusters form (as Fig. 2A,B). Here, a red factor is about to dissociate (arrow). 
{\bf (ii)} Green factors are phosphorylated; their affinity for pink beads is now higher than that of red factors. 
{\bf (iii)} A green factor has replaced a red one in the cluster due to higher-affinity binding.
{\bf (iv)} More green factors replace red ones in the cluster due to their higher affinity (see also Fig. 2B).
{\bf B.} Switching binding on/off by ``phosphorylation''/``dephosphorylation'' facilitates exchange with the soluble pool, as seen experimentally in photo-bleaching experiments~\cite{Brackley2017a}.
{\bf (i)} If factors exist permanently in a binding state, high local concentrations ensure they dissociate and rebind to the same cluster (as {\it 1}); consequently, there is little exchange with the soluble pool. 
{\bf (ii)} If factors switch between binding/non-binding states, they often exchange (here, the cluster loses {\it 2} and gains {\it 3}) and clusters can persist for hours as constituents exchange in seconds (as seen experimentally).
{\bf C.} Inheriting and erasing epigenetic states~\cite{Michieletto2017}.
{\bf (i)} A na\"ive string lacking ``epigenetic marks''. 
{\bf (ii)} Green and red ``bookmarks'' (e.g., factors related to active and inactive chromatin) bind to cognate beads to form green and red clusters (as Fig. 2A,B).
{\bf (iii)} Bookmarks now recruit epigenetic ``readers'' and ``writers'' (not shown) that ``mark'' histones in nearby beads (colored dots in the string). 
{\bf (iv)} Resulting ``epigenetic states'' and ``epigenetic domains'' persist through continued action of readers/writers. 
{\bf (v)} The system quickly restores marks when either marks or bookmarking factors are removed randomly (mimicking losses occurring during ``semi-conservative replication'' or ``mitosis''). 
{\bf (vi)} States are lost as the concentration of bookmarks becomes too dilute to maintain them (or if the genomic sequence binding the bookmark is excised, not shown~\cite{Michieletto2017}).
}
\end{minipage}

\vspace{\fill}

\begin{minipage}{\linewidth}
\centering
\includegraphics[width=0.5\textwidth]{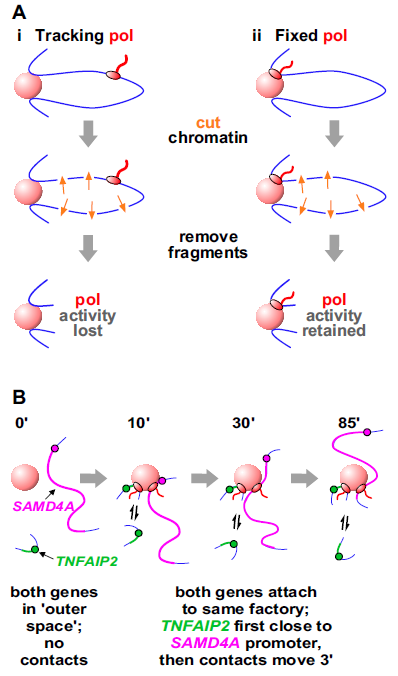}
\captionof{figure}
{\small
Two approaches showing that active polymerases cannot track like locomotives down templates (shown as loops tethered to a sphere; pol -- polymerase).
{\bf A.} The experiment involves nuclease digestion, removal of resulting fragments, and detection of $3$ markers -- remaining nascent RNA, transcribed sequence, and polymerizing activity (oval). 
{\bf (i)} If the active polymerase tracks, cutting chromatin should separate it from anchor points; when small fragments of chromatin are now removed, all $3$ markers will be lost. 
{\bf (ii)} If the polymerase anchors the loop, cutting chromatin and removing fragments will leave all $3$ markers; this is the result seen.
{\bf B.} Analysis of 3C contacts made between $11$-kbp {\it TNFAIP2} and $221$-kbp {\it SAMD4A}. Before adding TNF$\alpha$ ($0$ min) both genes are silent and not in contact. Ten minutes after adding TNF$\alpha$, both genes become active; {\it TNFAIP2} now often contacts the {\it SAMD4A} promoter (but not downstream segments). After $30$ min, {\it TNFAIP2} no longer contacts the {\it SAMD4A} promoter; instead, it contacts a point one-third into {\it SAMD4A}. After $60$ min, contacts shift two-thirds into {\it SAMD4A}, and by $85$ min they reach the terminus. These results are impossible to explain if polymerases track, but easily explained if the two active polymerases are immobilized in one factory. Then, after $10$ min, both genes attach to (and initiate in) this factory (giving promoter-promoter contacts). As it takes $85$ min to transcribe {\it SAMD4A} whilst {\it TNFAIP2} is transcribed in minutes, the short gene goes through successive transcription cycles by attaching to (and detaching from) the factory. Consequently, whenever {\it TNFAIP2} is transcribed, it will lie close to the point on {\it SAMD4A} that is being transcribed at that moment.
}
\end{minipage}

\vspace{\fill}

\begin{minipage}{\linewidth}
\centering
\includegraphics[width=\textwidth]{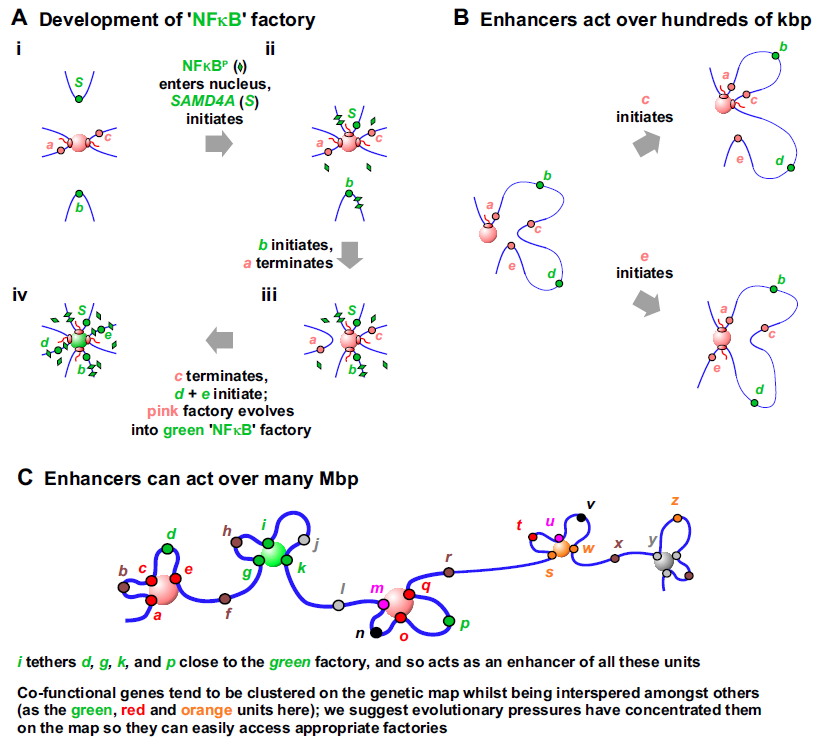}
\captionof{figure}
{\small 
{{The development of an ``NF$\kappa$B'' factory, and enhancer action, in a human cell.
{\bf A.} Development of an ``NF$\kappa$B'' factory on addition of TNF$\alpha$~\cite{Papantonis2012}. 
{\bf (i)} Before adding TNF$\alpha$. Promoters {\it a} and {\it c} have initiated in the pink factory; {\it S} ({\it SAMD4A}) and {\it b} may visit the factory, but they cannot initiate as the required transcription factor is absent. 3C shows that {\it S} and {\it b} rarely contact each other.
{\bf (ii)} $10$ min after adding TNF$\alpha$. NF$\kappa$B is now phosphorylated (NF$\kappa$BP), it entered the nucleus, and -- when {\it S} now visits the factory -- it initiates. {\it S} encodes many NF$\kappa$B-binding sites, and exchange of NF$\kappa$BP from these sites now creates a local concentration of the factor in/around the factory.
{\bf (iii)} {\it b} visits the factory and initiates. 3C shows {\it S} and {\it b} now often contact each other. Both genes encode NF$\kappa$B-binding sites, so the local concentration of the factor in/around the factory increases. 
{\bf (iv)} The pink factory develops into a (green) ``NF$\kappa$B'' factory specializing in transcribing green units as other green promoters initiate. 
{\bf B.} Enhancers can act over hundreds of kbp. Initially, {\it a}, {\it c}, and {\it e} were transcribed in the factory, but {\it c} and {\it e} have just terminated. {\it a} still tethers {\it c} and {\it e} close to the factory, and so both are likely to re-initiate. Consequently, {\it a} is an enhancer of {\it c} and {\it e}. As $\sim 10$ loops of $\sim 86$ kbp are typically anchored to one human factory, a can tether genes lying $\sim 860$ kbp away near the factory, and so enhance activity. 
{\bf C.} Enhancers can act over many Mbp. About $4$ Mbp of a human chromosomes are shown (again, only some of the $\sim 10$ loops/factory are shown). Transcription units {\it a} -- {\it z} tend to be transcribed in factories of the same color, except for purple ones that are promiscuous. Imagine {\it i} is transcribed often. Consequently, {\it d}, {\it g}, {\it k} and {\it p} will be tethered near the green factory so {\it i} acts as their enhancer (even though some lie $>1$ Mbp away). Co-functional genes (i.e., ones with promoters of similar color) also tend to be clustered on the genetic map, as shown here; we suggest this is the result of evolutionary pressures ensuring they can easily access appropriate factories. Note that green promoters are interspersed amongst pink ones, so it is possible this structure evolves into one where all green promoters are simultaneously transcribed in one green factory (while all pink promoters are transiently silent), and then into another structure where all pink promoters are transcribed in one pink factory (while all green promoters remain silent).}}}
\end{minipage}

\vspace{\fill}

\begin{minipage}{\linewidth}
\centering
\includegraphics[width=\textwidth]{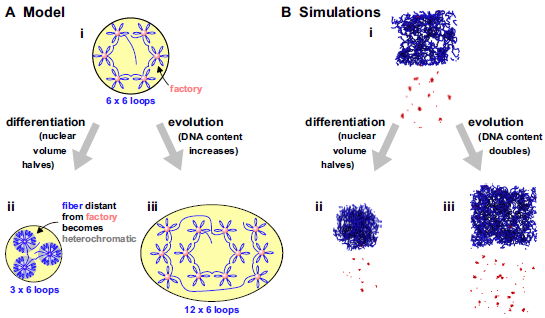}
\captionof{figure}
{\small 
A model and simulations indicating how promoter-factory distance can remain roughly constant despite changes in nuclear volume occurring during differentiation and evolution.
{\bf A.} Model~\cite{Faro2006}. 
{\bf (i)} All nucleoplasmic chromatin of a mouse ES cell is represented by one chromatin fiber organized into $6$ loops around $6$ factories.
{\bf (ii)} Differentiation into a cell with half the nucleoplasmic volume. Experimental data show the smaller nucleus has half the number of factories and active polymerases, but a similar factory diameter and density (i.e., number of factories per unit nucleoplasmic volume). As half the number of polymerases are active but the total amount of DNA is similar, our model requires the total number of loops should fall and contour length increase. Therefore, one might expect the volume of chromatin around each factory to increase. However, two factors probably combine to ensure it does not. First, polymer physics indicates that as loop length doubles, the radius of the volume occupied increases only $\sim 1.5$-fold~\cite{Bon2006}. Second, the fiber distant from a factory probably becomes heterochromatic and so more tightly packed (grey zone). Consequently, increased loop length has little effect on factory density. In other words, the system self-regulates so the average gene remains just as far away from a factory despite the volume change.
{\bf (iii)} Changes occurring during evolution as DNA content increases $2$-fold (original data involved comparison of a mouse ES cell and a newt cell with $10$-fold more DNA). Factory diameter and density remain constant, as nucleoplasmic volume and total number of active polymerases increase. As there is more DNA and more polymerases are active, we suggest loop contour-length remains constant; the system again self-regulates.
{\bf B.} Snapshots from $3$ Brownian-dynamics simulations consistent with the model in (A). Simulations (details in Supplementary Note 6) involve a string (``chromatin'' fiber) of blue beads (each representing $3$ kbp) diffusing in a cube as red spheres (``factors''/``polymerases'') bind reversibly to cognate beads spaced every $90$ kbp along the string (also shown blue, interaction energy and range -- $7.1$ kBT and $54$ nm). Upper and lower panels show images of all beads plus ``factors'', or just ``factors'', in the cube at the end.
{\bf (i)} Stem cell ($15$-Mbp fiber, $100$ factors, $1.5$ $\mu$m side cube). Bound red beads spontaneously cluster (as Fig. 2A).
{\bf (ii)} During ``differentiation'', the same amount of chromatin is confined in half the volume ($1.2$ $\mu$m side cube), and there are half the number of factors and binding beads (reflecting silencing of half binding sites). The number of red beads/cluster, cluster density, and cluster diameter are as (i), but cluster number halves.
{\bf (iii)} During ``evolution'' to a cell with twice the DNA, ``fiber'' length doubles to $30$ Mbp, but ``chromatin'' and ``factor'' density remain constant ($1.89$ $\mu$m side cube; $200$ ``factors''). Cluster number doubles, but the number of red beads/cluster, cluster density, and cluster diameter are again as (i).
}
\end{minipage}

\end{document}